\documentclass[a4paper]{jpconf}
\usepackage{graphicx}
\usepackage{multirow}
\usepackage{amsmath}
\usepackage[psamsfonts]{amssymb}
\usepackage{amsxtra}
\usepackage{threeparttable}
\usepackage{url}
\usepackage[hidelinks]{hyperref}

\begin{document}
\title{Evaluation of Error and Correlation-Based Loss Functions 
For Multitask Learning Dimensional Speech Emotion Recognition}

\author{Bagus Tris Atmaja$^{1,2}$, Masato Akagi$^{2}$}

\address{$^{1}$Sepuluh Nopember Institute of Technology, Surabaya, Indonesia\\
$^{2}$Japan Advanced Institute of Science and Technology, Nomi, Japan}

\ead{bagus@ep.its.ac.id, akagi@jaist.ac.jp}

\begin{abstract}
The choice of a loss function is a critical part in machine learning. This
paper evaluates two different loss functions commonly used in regression-task
dimensional speech emotion recognition --- error-based and correlation-based
loss functions. We found that using correlation-based loss function with
concordance correlation coefficient (CCC) loss resulted in better performance
than error-based loss functions with mean squared error (MSE) and mean absolute
error (MAE). The evaluations were measured in averaged CCC among three
emotional attributes. The results are consistent with two input feature sets
and two datasets.  The scatter plots of test prediction by those two loss
functions also confirmed the results measured by CCC scores.
\end{abstract}

\section{Introduction}
% DSER
Dimensional emotion recognition is scientifically more challenging than
categorical emotion recognition. In dimensional emotion recognition, the goal
is to predict the continuous degree of emotional attributes, while in
categorical emotion recognition, the task is to predict emotion category of
speakers, whether they are angry, happy, sad, fearful, disgusting, or
surprised.  While the practical application of this dimensional emotion
recognition is not clear yet, Russel \cite{Russell1979} argued that
categorical emotion can be derived from two-space dimensional emotion, i.e.,
valence (positive or negative) and arousal (high or low).

% Multitask learning
In the dimensional emotion model, several models have been introduced by
psychological researchers, including 2D, 3D, and 4D models. In the 3D model,
either dominance (power control) or liking is used as the third attribute. In
the 4D model, either expectancy or unpredictability is used, such as in
\cite{Moore2014word} and \cite{Fontaine2017}. This research used 3D emotions
with valence, arousal, and dominance (VAD) model, as suggested in
\cite{Bakker2014}.

% Loss function
In the 3D emotion model, the emotion recognizer system (classifier) needs to
predict three emotional attributes. This task is often performed by
simultaneous or jointly learning prediction of VAD. This simultaneous learning
technique is known as multitask learning (MTL). Compared to single-task
learning (STL), MTL attempts to optimize three parameters at the same time,
while STL only attempts to optimize one parameter (either valence, arousal, or
dominance).  To train the model over those three attributes, the choice of loss
function for the MTL is vital for the performance of the system. The
traditional regression task used mean squared error (MSE) as both loss function
and evaluation metric. Dimensional emotion recognition, as a regression task,
conventionally follow that rule by applying MSE for both loss and evaluation
metric. Additionally, mean absolute error (MAE) is another metric to train
and evaluate regression task. MAE loss function is less used since it only can
be compared on the same-scale data.

% Argued no direct comparison MSE vs. CCC
Recently, affective computing researchers argued that using correlation-based
metric to evaluate the performance of dimensional emotion recognition is more
appropriate than calculating its errors \cite{Ringeval2015a, Han2017a,
Schmitt2019}.  The concordance correlation coefficient (CCC)
\cite{lawrence1989concordance} is often used to measure the performance of
dimensional emotion recognition since it takes the bias into Pearson's
correlation coefficient (CC). Hence, we hypothesized that using CCC loss
($1-CCC$) is more relevant than using MSE and MAE as loss functions for MTL
dimensional speech emotion recognition. This paper aims to evaluate this
hypothesis.

% Contribution of paper
To the best of our knowledge, there is no experimental research reporting
direct comparison of the impact of using error- vs. correlation-based loss
function for dimensional speech emotion recognition.  The mathematical
foundation of this issue has been thoroughly explained in \cite{Pandit2019}.
Some authors used MSE loss, such as in \cite{parthasarathy2017jointly,
Abdelwahab2018} while the others used CCC loss, such as in \cite{Schmitt2019,
Atmaja2020}. Both groups reported the performance of the evaluated method
using CCC.  We choose speech emotion recognition as our task since the target
application is speech-based applications like voice assistant and call center
service.

\section{Problem Statement}
We focused our work to evaluate which loss function performs better on
multitask learning dimensional emotion recognition: MSE, MAE, or CCC loss. To
achieve this goal, we used two different datasets and two different acoustic
feature sets. We expected consistent results across four scenarios or parts (2
datasets $\times$ 2 feature sets).  The same experiment condition (i.e., the
same architecture with the same parameters) is used to evaluate both
error-based and correlation-based loss functions in four scenarios. The
evaluation is measured by CCC scores.  The averaged CCC score among three
emotion dimensions is used to evaluate the overall performance of evaluated loss
function on each scenario.

\section{Evaluation Methods}
% This section presents the research methods in sequential: data and
% feature sets, MSE-based loss function, CCC-based loss function, and
% architecture of dimensional speech emotion recognition.

\subsection{Data and Feature Sets}
Two datasets and two acoustic feature sets were used to evaluate three different
loss functions (CCC loss, MSE, MAE).

\noindent\textbf{Datasets:} IEMOCAP and MSP-IMPROV datasets are utilized to
evaluate error and correlation-based loss functions. Among many modalilites
provided by both datasets, only speech data is used to extract acoustic feature
sets. The first dataset consists of 10039 utterances while the second consists
of 8438 utterances. For both datasets, only dimensional labels are used, i.e.,
valence, arousal, and dominance, in the range [1, 5].  We scaled those labels
into the range [-1, 1] following the work in \cite{parthasarathy2017jointly}
when fed it into deep learning-based dimensional speech emotion recognition
system. This new scale is more readable because valence, arousal, and dominance
are from negative to positive scale by their definitions. The detail of IEMOCAP
dataset is given in \cite{Busso2008}, while for MSP-IMPROV dataset is available
in \cite{busso2016msp}. All scenarios in the two datasets are performed in
speaker-independent configuration for test data, i.e., the last one session is
left out for test partition (LOSO, leave one session out). For IEMOCAP data,
the number of utterances in training partition is 7869 utterances, and the rest
2170 utterances (session fife) are used for the test partition. On the
MSP-IMPROV dataset, 6816 utterances are used for the training partition, and
the rest 1622 utterances (session six) are used for the test partition. On both
training partitions, 20\% of data is used for validation (development).

\noindent\textbf{Acoustic Features:} High-level statistical function (HSF) of
two feature sets are used. The first is HSF from Geneva minimalistic acoustic
and parameter set (GeMAPS), as described in \cite{Eyben}. The second is HSF
from pyAudioAnalysis (pAA) \cite{Giannakopoulos2015}. Note that the definition
of HSF referred here is only mean and standard deviation (Mean+Std) from
low-level descriptors (LLD) listed in both feature sets. GeMAPS feature set
consists of 23 LLDs, while pAA contains 34 LLDs.  A list of LLDs in those two
feature sets is presented in table \ref{tab:feature}.  The use of Mean+Std in
this research follows the finding in \cite{Schmitt2018}.  Additionally, we
implement the Mean+Std of LLDs from pAA to observe its difference from GeMAPS.

\begin{table}[htpb]
\begin{center}
\caption{Acoustic features used to evaluate the loss functions (only Mean+Std 
of those LLDs are used as input features).}
\label{tab:feature}
\begin{tabular}{l p{11cm}}
\br
Feature set    &   LLDs \\
\mr
GeMAPS & intensity, alpha ratio, Hammarberg index, spectral slope 0-500 Hz,
spectral slope 500-1500 Hz, spectral flux, 4 MFCCs, $f_o$, jitter, shimmer,
Harmonics-to-Noise Ratio (HNR), harmonic difference H1-H2, harmonic difference
H1-A3, F1, F1 bandwidth, F1 amplitude, F2, F2 amplitude, F3, and F3 amplitude.
\\ 
    \mr
pAA & zero crossing rate, energy, entropy of energy, spectral centroid,
spectral spread, spectral entropy, spectra flux, spectral roll-off, 13 MFCCs,
12 chroma vectors, chroma deviation.\\
    \br
\end{tabular}
\end{center}
\end{table}

\subsection{Error-based Loss Function}
A mean squared error to measure the deviation between predicted emotion degree
$x$ and gold-standard label $y$ is given by 
\begin{equation}
MSE ={\frac {1}{n}}\sum _{i=1}^{n}(x_{i}-{y_{i}})^{2},
\end{equation}
where $n$ is the number of measurement (calculated per batch size). For three
emotion dimensions, the total MSE is the sum of MSE from valence, arousal, and
dominance: 
\begin{equation}
MSE_T = MSE_V + MSE_A + MSE_D.
\end{equation}
Following the work of \cite{parthasarathy2017jointly}, we added weighting
factors for valence and arousal. Hence, the total MSE became, 
\begin{equation}
MSE_T = \alpha MSE_V + \beta MSE_A + (1-\alpha-\beta) MSE_D,
\end{equation}
where $\alpha$ and $\beta$ are weighting factors for valence and arousal.  The
weighting factor for dominance is obtained by subtracting 1 with those two
variables.

Similarly, mean absolute errors as an objective function can be formulated as
follows, 
\begin{gather}
MAE ={\frac {1}{n}}\sum _{i=1}^{n}|x_{i}-{y_{i}}|, \\
MAE_T = MAE_V + MAE_A + MAE_D, \\
MAE_T = \alpha MAE_V + \beta MAE_A + (1-\alpha-\beta) MAE_D.
\end{gather}
Although the use of MAE as a loss function is less common in machine learning
due to a scale-dependent accuracy measure, we evaluated it since both dataset
labels are in the same scale ([1, 5] standardized to [-1, 1]). This additional
loss function might support the comparison of error-based to correlation-based
loss functions.

\subsection{Correlation-based Loss Function}
CCC is a common metric in dimensional emotion recognition to measure the
agreement between the true emotion dimension with predicted emotion degree. If
the predictions shifted in value, the score is penalized in proportion to
deviation \cite{Ringeval2015a}.  Hence, CCC is more reliable than Pearson
correlation, MAE and MSE to evaluate the performance of dimensional speech
emotion recognition. CCC is formulated as 
\begin{equation}
    CCC = \dfrac {2\rho_{xy} \sigma_{x} \sigma_{y}}
        {{\sigma_{x}^2}+\sigma_{y}^2 + (\mu_x - \mu_y)^2},
\end{equation}

\noindent
where $\rho_{xy}$ is the Pearson coefficient correlation between $x$ and $y$,
$\sigma$ is the standard deviation, and $\mu$ is a mean value.  This CCC is
based on Lin's calculation \cite{lawrence1989concordance}.  The range of CCC is
from $-1$ (perfect disagreement) to $1$ (perfect agreement).  Therefore, the
CCC loss function (CCCL) to maximize the agreement between true value and
predicted emotion can be defined as 
\begin{equation}
    CCCL = 1 - CCC.
\end{equation}
Similar to multitask learning in MSE, we accommodate the loss functions from
arousal ($CCCL_{V}$), valence ($CCCL_{A}$), and dominance ($CCCL_{D}$).  The
$CCCL_{T}$ is a combination of these three CCC loss functions:
\begin{equation}
    CCCL_{T} = \alpha ~ CCCL_{V} + \beta ~ CCCL_{A} + (1-\alpha-\beta) ~ CCCL_{D},
\label{eq:ccc}
\end{equation}
where $\alpha$ are $\beta$ are the weighting factors for each emotion dimension
loss function. The same weighing factors are used for MSE, MAE, and CCC losses,
i.e., $\alpha=0.1$ and $\beta=0.5$ for IEMOCAP dataset, and $\alpha=0.3$ and
$\beta=0.6$ for MSP-IMPROV dataset. Those weighting factors are obtained via
linear search in range [0.0, 1.0] with $0.1$ increment dependently each other.

\subsection{Architecture of Dimensional Speech Emotion Recognition}
\begin{figure}
    \centering
        \includegraphics[width=.6\textwidth]{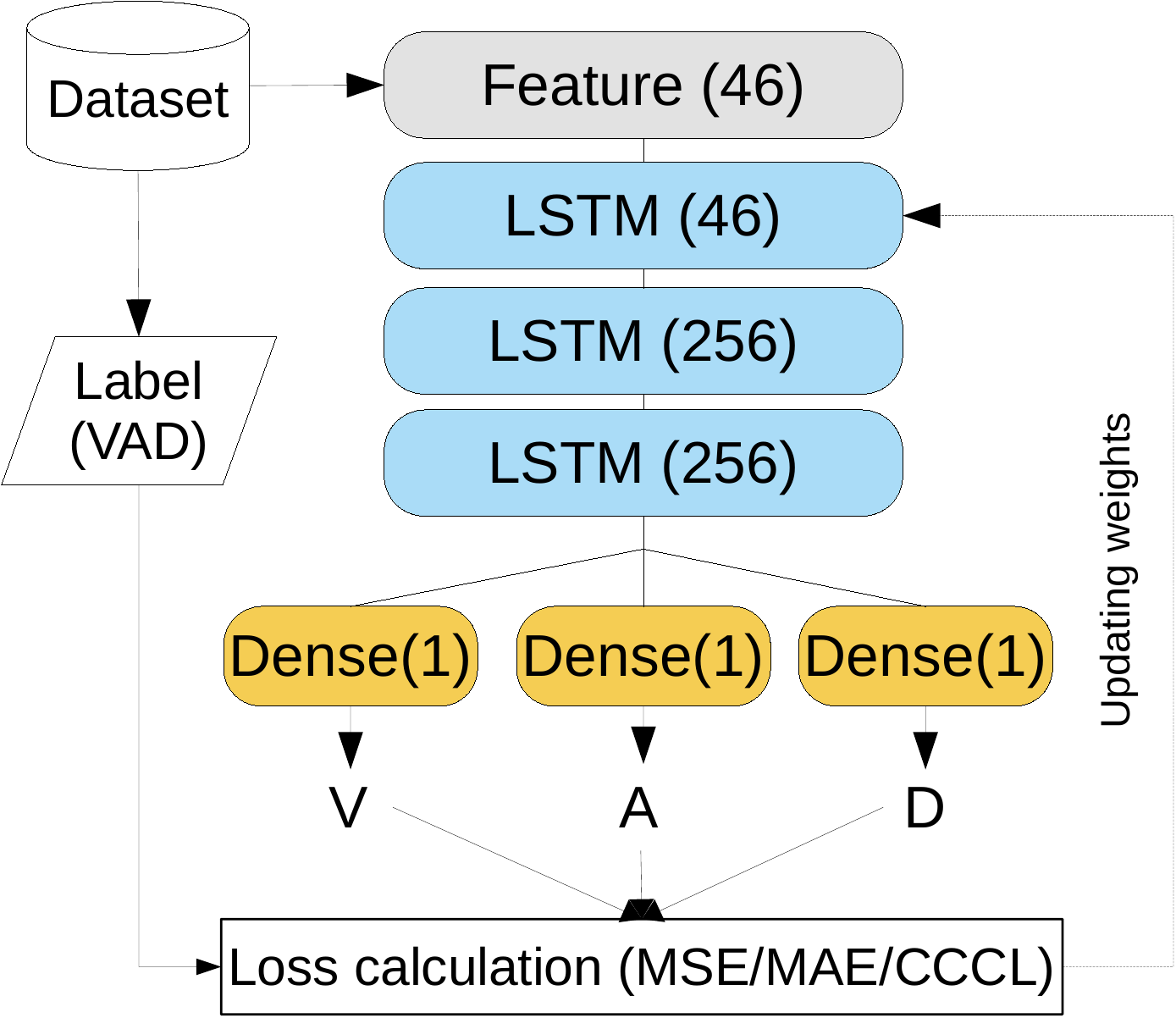}
    \caption{Architecture of dimensional speeh emotion recognition system to 
    evaluate loss functions.}
    \label{fig:dser}
\end{figure}

We used deep learning-based architecture to evaluate all loss functions, i.e.,
three layers of stacked LSTM networks \cite{Hochreiter1997}.  For the input,
either 46 HSFs from GeMAPS or 68 HSFs from pAA are fed into the network. A
batch normalization layer is performed to speed up the computation process
\cite{Ioffe2015}. Three LSTM layers are stacked; the first two layers return
all sequences while the last LSTM layer returns final outputs only. A dense
network with 64 nodes is coupled after the last LSTM layer. Three dense layers
with one unit each ended the network to predict the degree of valence, arousal,
and dominance with \verb|tanh| activation function to bound the output in range
[-1, 1]. Either MSE, MAE, or CCC loss is used as the loss function with RMSprop
optimizer \cite{tieleman2012lecture}.  The architecture of this dimensional
speech emotion recognition is shown in Fig. \ref{fig:dser}.

As an additional analysis tool, we used scatter plots of predicted valence and
arousal degrees compared to the gold-standard labels. These plots will show how
similar or different between labels and predicted degrees.  This similarity
between labels and predictions can be used to confirm obtained CCC scores by
different loss functions. 

The implementation of the evaluation methods is available in the following
repository, \url{https://github.com/bagustris/ccc_mse_ser}.  The LSTM-based
dimensional speech emotion recognition is implemented using Keras toolkit
\cite{chollet2015keras} with TensorFlow backend \cite{abadi2016tensorflow}.

\section{Results and Discussion}
\subsection{Performance of Evaluated Loss Functions}
The main issue of this research is to find which loss function works
better for multitask learning dimensional emotion recognition. The following
results report evaluation of three loss functions in two datasets and two
features sets.
%For the results, two metrics are presented, CCC and MSE. However, since 
%CCC is being used as the gold-standard measure of dimensional emotion recognition, 
%only CCC is counted for the evaluation. 
%Particularly, the average CCC score from valence, arousal, and dominance 
%is used to determine which loss function performed better.

Table \ref{tab:result} shows the result of using different loss function in the
same dataset using the same network architecture. We can divide the results
into four parts: two datasets with two different feature sets for each dataset.
The first part is the IEMOCAP dataset with HSF of GeMAPS as the input feature.
Using CCC loss, the obtained CCC score for each dimension is higher than the
obtained scores using MSE and MAE losses. The resulted average CCC scores were 
0.304 for MAE, 0.310 for MSE, and 0.400 for CCCL. Clearly, it is shown that
CCCL obtained better performance than MSE and MAE  in this part and continued
to other three parts.

We evaluated HSFs from pAA on the second part of the table. Although the
feature set is not designed specifically for an affective application, however,
it showed a similar performance to the result obtained by affective-designed
GeMAPS feature set. In this IEMOCAP dataset, HSF of pAA even performed
marginally better than HSFs of GeMAPS for both CCC and MSE losses. The
comparison of performance between MSE, MAE, and CCCL is similar to those were
obtained by GeMAPS, with the average CCC score of 0.333, 0.344, and 0.410 for
MSE, MAE,  and CCC loss, respectively.

Moving to the MSP-IMPROV dataset, a similar trend was observed. On the third
part with MSP-IMPROV and GeMAPS feature set, CCCL obtained the averaged CCC
score of 0.363 compared to MAE with 0.323 and MSE with 0.327. Finally, on the
fourth part with MSP-IMPROV and pAA feature set, the CCCL obtained 0.34 of the
average CCC score, while MSE and MAE obtained 0.305 and 0.324.

The overall results above suggest that, in terms of CCC, CCC loss is better
than MSE and MAE for multitask learning dimensional emotion recognition. Four
scenarios with CCC loss function obtained higher scores than other four
scenarios with MSE and MAE, on both individual emotion dimensions scores (CCC
of valence, arousal, and dominance) and on the averaged score. We extend the
discussion to the results obtained by MSE scores and different feature sets.

We found that the averaged MSE scores across data and feature sets are almost
identical (last column in Table \ref{tab:result}). If so, the MSE metrics might
be more stable to generalize the model generated by dimensional speech emotion
recognition system across different datasets. However, this consistent error
for the generalization of dimensional speech emotion recognition system needs
to be investigated with other datasets in different scale of labels.  Another
possible cause for the consistent error is the small range of the output, i.e.,
0-1 scale after squared by the MSE.

\begin{table}[!htbp]
    \centering
    \caption{Evaluation results of evaluated loss functions on IEMOCAP and MSP-IMPROV datasets.}
    \label{tab:result}
    \begin{tabular}{l l  c c c c}
    \br
    Feature & Loss    &   \multicolumn{4}{c}{CCC}   \\
        &   &   V   &   A   &   D   & Mean   \\
    \mr
    \multicolumn{6}{c}{IEMOCAP} \\
    GeMAPS  & MSE & 0.121 & 0.451 & 0.358 & 0.310 \\
            & MAE & 0.145	& 0.431	& 0.337	& 0.304 \\
            & CCCL & 0.192 & 0.553 & 0.456 & \textbf{0.400} \\
    pAA     & MSE & 0.093 & 0.522 & 0.383 & 0.333 \\
            & MAE & 0.110	& 0.539	& 0.381	& 0.344 \\
            & CCCL & 0.183 & 0.577 & 0.444 & \textbf{0.401} \\
    \multicolumn{6}{c}{MSP-IMPROV} \\
    GeMAPS  & MSE & 0.138 & 0.492 & 0.353 & 0.327 \\
            & MAE & 0.187	& 0.466	& 0.317	& 0.323 \\
            & CCCL & 0.204 & 0.525 & 0.361 & \textbf{0.363} \\
    pAA     & MSE & 0.122 & 0.475 & 0.319 & 0.305 \\
            & MAE & 0.148	& 0.479	& 0.343	& 0.324 \\
            & CCCL & 0.150 & 0.496 & 0.374 & \textbf{0.340} \\
    \br
    \end{tabular}
\end{table}

On the use of different feature sets, we observed no significant differences
between results obtained by HSF of affective-designed GeMAPS and
general-purpose pAA feature sets. The results obtained by those two feature
sets are quite similar for the same loss function. Not only on CCC scores, but
the similarity of performance is also observed on MSE scores. This results can
be viewed as the generalization from the previous research \cite{Schmitt2019}
that not only Mean+Std of GeMAPS useful for dimensional emotion recognition but
also Mean+Std of other feature sets, in this case, pAA feature set.

\subsection{Scatter Plot of Predicted Emotion Degrees}
\begin{figure}[htbp]
    \centering
        \includegraphics[width=0.55\textwidth]{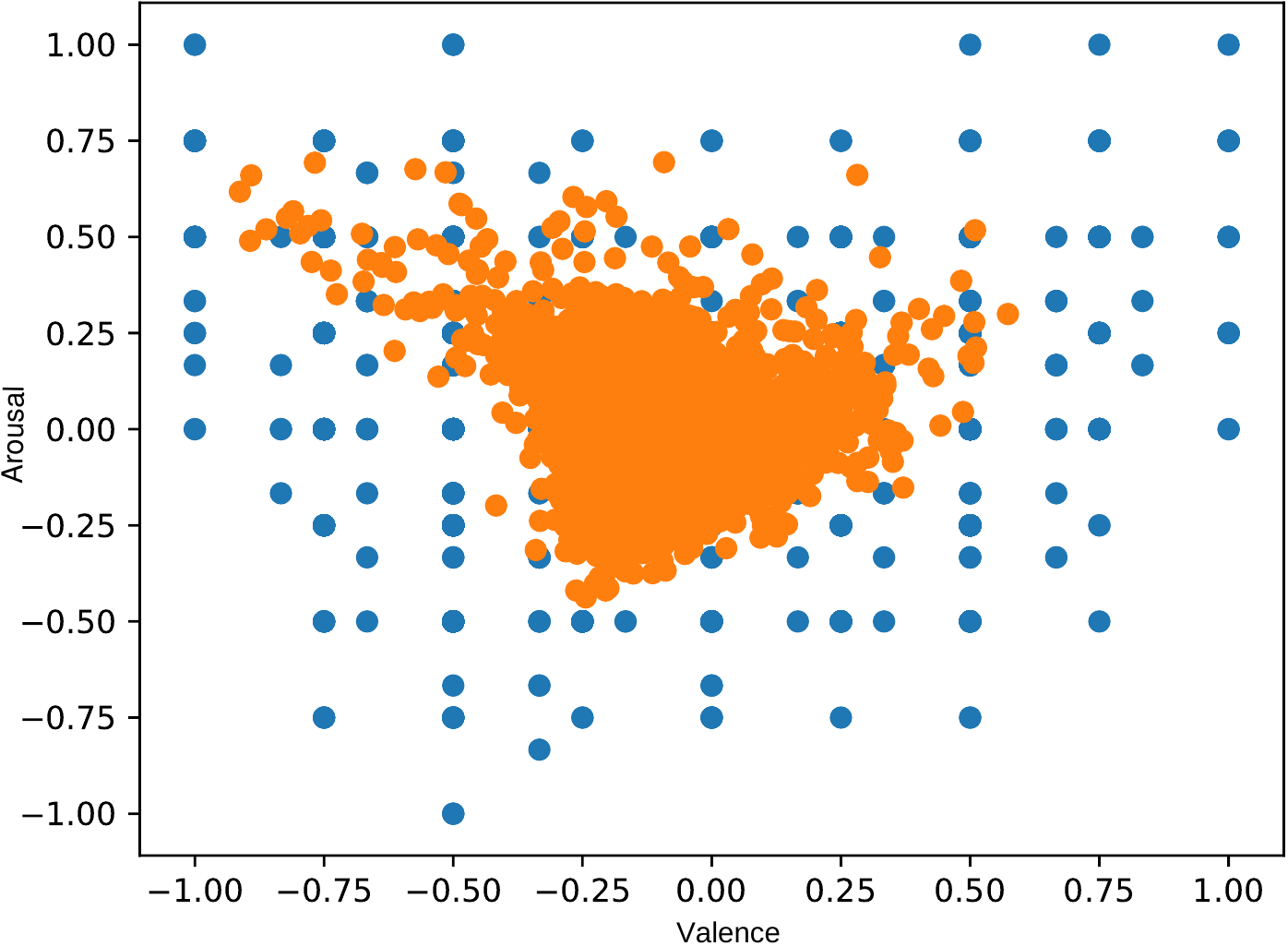} 

        \vspace{.5cm}
        
        \includegraphics[width=0.55\textwidth]{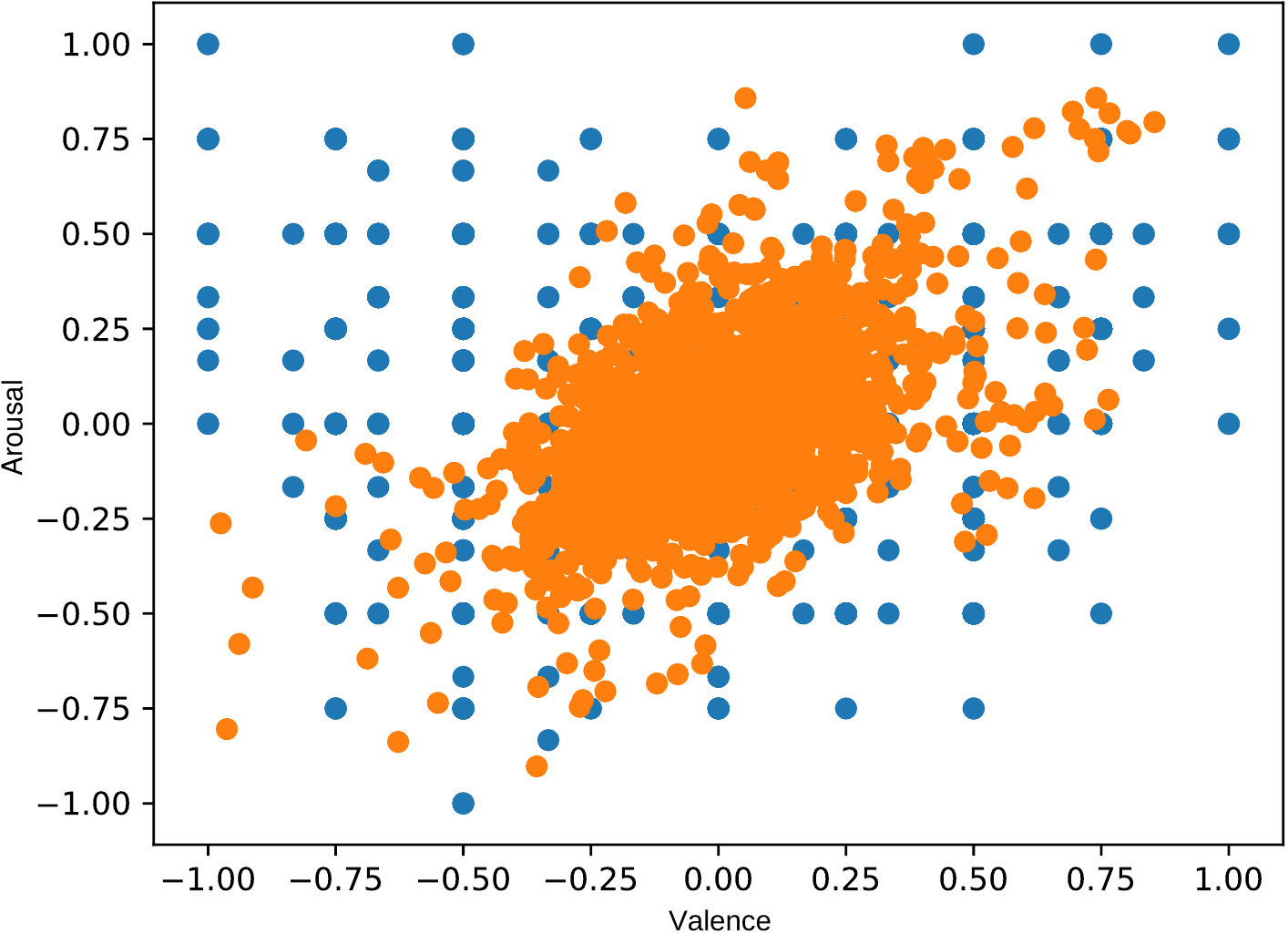}

        \vspace{.5cm}

        \includegraphics[width=0.55\textwidth]{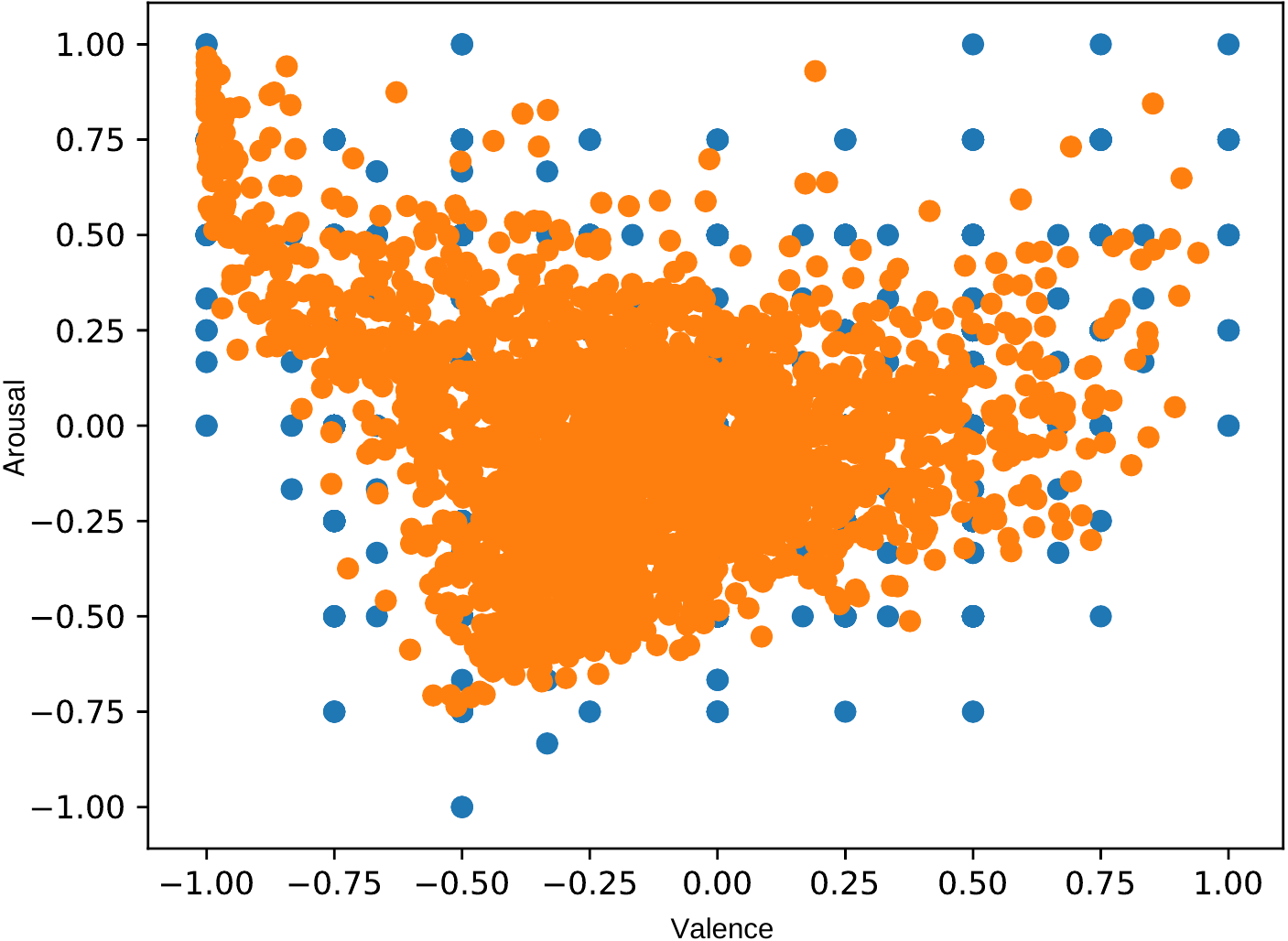}
    \caption{Scatter plot of valence and arousal dimensions (in CCC [-1, 1])
           on test partition
           from IEMOCAP dataset with GeMAPS features; top: MSE loss, middle;
           MAE loss, bottom: CCC loss; blue: 
           gold-standard label, orange: predicted degree).}
    \label{fig:scatter-ccc}
\end{figure}

We showed the scatter plots of prediction by and MSE, MAE, and CCC loss in Fig.
\ref{fig:scatter-ccc}. In these cases, the plots showed the prediction from
GeMAPS test partition. From both plots, we can infer that the prediction from
CCC loss is more similar to gold-standard labels than the prediction from MSE
and MAE. This result confirms the obtained CCC scores from valence, arousal, and
dominance and its average. Although we only showed the result from IEMOCAP with
GeMAPS feature, the plots are consistent with other parts.  Note in these
scatter plots that a single dot may represent more than one label (overlapped),
since there is a possibility to have the same labels (score of valence and
arousal) for several utterances.

Comparing the shape of three predictions shows that CCC
loss predicts better than MSE and MAE. The calculation of CCC takes account of
the shifted values of prediction into concordance correlation calculation while
MSE and MAE only count their errors. If the concordance line represents the
true valence-arousal label in 2D space (like in \ref{fig:scatter-ccc}), CCC
loss minimizes the variation of trained labels to this line. It can be
concluded from both metric and visualization that CCC loss performs better
than MSE and MAE. 

%   \begin{figure}[!htpb]
%       \centering
%           \includegraphics[width=0.45\textwidth]{fig/scatter_gemaps_mse.pdf}
%       \caption{Scatter plot of valence and arousal dimensions (in CCC [-1, 1])
%              on test partition from IEMOCAP dataset with GeMAPS feature and MSE
%              loss (blue: gold-standard label, orange: predicted degree).}
%       \label{fig:scatter-mse}
%   \end{figure}

%   \begin{figure}[!htpb]
%     \centering
%         \includegraphics[width=0.45\textwidth]{fig/scatter_gemaps_mae.pdf}
%     \caption{Scatter plot of valence and arousal dimensions (in CCC [-1, 1]) on
%            test partition from IEMOCAP dataset with GeMAPS feature and MAE
%            loss (blue: gold-standard label, orange: predicted degree).}
%     \label{fig:scatter-mse}
%     \end{figure}

\section{Conclusions}
This paper reported an evaluation of different loss functions for multitask
learning dimensional emotion recognition. The result shows that the use of CCC
loss obtained better performance than the MAE and MSE in CCC scores measure
across four scenarios.  We are confident that this result is universal since we
use two different datasets and two different feature sets that resulting
consistent results. The process is also straightforward, CCC loss as the loss
function with CCC scores as evaluation metrics. These results are also
supported by scatter plots of the valence-arousal prediction from both losses
compared to the gold-standard labels.

% On the other side, we also found that the use of MSE as a metric resulting 
% a more consistent errors across datasets and scenarios. 
%This finding may be 
%interpreted as, the error obtained by the same system remains similar 
%(using the same feature and loss function) while the measured performance 
%(correlation/correctness) is different. 
Further study to investigate the relationship between error and (concordance)
correlation from both theoretical and practical approaches may improve our
understanding on it.  Although it is suggested to use CCC as the main metric
for dimensional emotion recognition, additional metrics such as MSE and RMSE
might be useful to accompany CCC measure for tracking the pattern of the
performance across different datasets, feature sets, and methods, particularly
in dimensional emotion recognition. \\

\section*{References}
\bibliographystyle{iopart-num}
\bibliography{ccc_mse}

\end{document}